\documentclass[sigconf]{acmart}

\usepackage{hyperref}
\usepackage{cleveref}

\usepackage{dirtytalk} 
\usepackage{tabularx}

\AtBeginDocument{%
  }

\setcopyright{acmlicensed}
\copyrightyear{2024}
\acmYear{2024}
\acmDOI{10.1145/3674805.3690754}

\acmConference[ESEM'24]{International Symposium on Empirical Software Engineering and Measurement}{October 20--25,
  2024}{Barcelona, ES}

\acmISBN{979-8-4007-1047-6/24/10}

\begin{document}

\title{Crossover Designs in Software Engineering Experiments: Review of the State of Analysis}

\author{Julian Frattini}
\affiliation{
  \institution{Blekinge Institute of Technology}
  \city{Karlskrona}
  \country{Sweden}}
\email{julian.frattini@bth.se}

\author{Davide Fucci}
\affiliation{
  \institution{Blekinge Institute of Technology}
  \city{Karlskrona}
  \country{Sweden}}
\email{davide.fucci@bth.se}

\author{Sira Vegas}
\affiliation{
  \institution{Universidad Polit\'ecnica de Madrid}
  \city{Madrid}
  \country{Spain}}
\email{svegas@fi.upm.es}
\renewcommand{\shortauthors}{Frattini et al.}

\begin{abstract}
    Experimentation is an essential method for causal inference in any empirical discipline.
    Crossover-design experiments are common in Software Engineering (SE) research.
    In these, subjects apply more than one treatment in different orders.
    This design increases the amount of obtained data and deals with subject variability but introduces threats to internal validity like the learning and carryover effect.
    Vegas et al. reviewed the state of practice for crossover designs in SE research and provided guidelines on how to address its threats during data analysis while still harnessing its benefits.
    In this paper, we reflect on the impact of these guidelines and review the state of analysis of crossover-design experiments in SE publications between 2015 and March 2024.
    To this end, by conducting a forward snowballing of the guidelines, we survey 136 publications reporting 67 crossover-design experiments and evaluate their data analysis against the provided guidelines.
    The results show that the validity of data analyses has improved compared to the original state of analysis.
    Still, despite the explicit guidelines, only 29.5\% of all threats to validity were addressed properly.
    While the maturation and the optimal sequence threats are properly addressed in 35.8\% and 38.8\% of all studies in our sample respectively, the carryover threat is only modeled in about 3\% of the observed cases.
    The lack of adherence to the analysis guidelines threatens the validity of the conclusions drawn from crossover-design experiments.
\end{abstract}




\maketitle

\section{Introduction}
\label{sec:intro}

Experimentation is an important method to infer causal relationships in any empirical research discipline~\cite{wohlin2012experimentation}.
The design of an experiment, i.e., how levels of the main factor are assigned to subjects, has a critical impact on the validity of its conclusions.
One possible design, the \textit{crossover} design, has the advantage of increasing the number of data points obtained for the same amount of subjects involved in an experiment but is often critiqued for introducing threats to validity due to its difficult analysis~\cite{kitchenham2003case}.
To mitigate these threats, Vegas et al.~\cite{vegas2015crossover} provided a thorough explanation and guidelines for the analysis of crossover-design experiments.

In this paper, we reflect on the impact of the guidelines by Vegas et al.~\cite{vegas2015crossover} by answering the following research question (RQ): How do SE experiments that utilize a crossover design based on guidelines by Vegas et al. analyze their data?
To this end, we conduct a forward snowballing of primary studies citing the guidelines~\cite{vegas2015crossover} and assess how they deal with the threats to validity for which this experimental design is often critiqued~\cite{kitchenham2003case}.

Our contribution is three-fold:

\begin{enumerate}
    \item We reproduce and archive the data analysis from the original guidelines~\cite{vegas2015crossover} which were previously unavailable.
    \item We show gaps in SE literature on analyzing data from cross\-over-design experiments.
    \item We reflect on the impact of the original guidelines~\cite{vegas2015crossover} on the landscape of SE experimentation since its publication.
\end{enumerate}


The rest of the paper is structured as follows:
\Cref{sec:related} explains different experimental designs and reviews literature studies with a similar purpose. 
\Cref{sec:method} reports the applied method, \Cref{sec:results} presents the results, and \Cref{sec:discussion} discusses their implications.
\Cref{sec:limit} acknowledges threats to validity and outlines future work, before we conclude the paper in \Cref{sec:conclusion}.

\subsection*{Data Availability Statement}

All data, protocols, material, figures, and scripts generated and used during this study are publicly available~\cite{replicationpackage}.

\section{Background and Related Work}
\label{sec:related}

During an experiment, researchers apply one or more levels of a main factor (i.e., treatments) to experimental subjects and observe one or more dependent variables assumed to be impacted by the factor.
Observing a significant difference in the distribution of a dependent variable for different treatments allows the inference that this factor has a causal relationship with the independent variable.
These significant differences are typically determined by selecting and applying an appropriate null-hypothesis significance test (NHST)~\cite{wohlin2012experimentation}, like the T-Test, Mann-Whitney U test, or ANOVA.
Experimentation has been used in software engineering (SE) to determine the effect of different tools~\cite{sandobalin2020effectiveness}, methods~\cite{trkman2019impact}, but also demographic factors~\cite{bogner2023restful} on SE tasks.
\Cref{sec:related:design} presents the concept and challenges of experimental design.
\Cref{sec:related:previous} provides an overview of previous studies with a similar purpose to ours.

\subsection{Experimental Design}
\label{sec:related:design}

An important decision when designing an experiment is whether each subject is administered only one or multiple levels of the main factor.
The former is referred to as an \textit{independent measures} (or between-subjects) design, where subjects are split into groups and each group applies only one treatment, the latter is referred to as a \textit{repeated measures} (or within-subjects) design.
The benefit of a repeated measures design is two-fold.
Firstly, the same amount of experimental subjects produce more data points than in an independent measure design.
Secondly, comparing relative rather than absolute values of each response variable deals with the threat of subject variability, i.e., that tasks performed in SE experiments are often influenced by differences between human subjects~\cite{pickard1998combining} which are difficult to quantify.
On the other hand, such a design introduces new threats to validity.
Vegas et al. identified three additional types of threats in their guidelines~\cite{vegas2015crossover}:

\begin{enumerate}
    \item \textit{Maturation/exhaustion:} Participants may perform better or worse in subsequent observations due to a learning or exhaustion effect.
    \item \textit{Optimal sequence:} Some sequences in which the treatments are administered may be favorable over others.
    \item \textit{Carryover}: The effect of an administered treatment might carry over to a subsequent period.
\end{enumerate}

The crossover design, a special form of the repeated-measures design where participants receive the treatments in different sequences~\cite{vegas2015crossover}, counterbalances the threats by dispersing their impact on the response variable evenly.
Still, the threats (including the threat of subject variability) affect the observations of the response variable but are largely ignored in most SE papers~\cite{vegas2015crossover}.
In medical research, the carryover effect is often addressed via a washout period~\cite{woods1989two}, i.e., a period between the experimental periods long enough for the medical compound to exit the subject's system.
In SE research, a washout period is often not feasible since it would require participants to unlearn techniques or tools they were subjected to~\cite{kitchenham2003case}.
Still, many SE papers even fail to acknowledge the carryover threat~\cite{vegas2015crossover}.

Consequently, Vegas et al. recommended abandoning simple NHSTs that only test whether different treatments change the distribution of the response variable and rather adopting linear mixed models (LMMs) to analyze the data from crossover-design experiments.
An LMM can involve additional factors with an influence on the response variable, which (1) isolates the true effect of the main factor in question and (2) provides insight into whether the threats to validity apply in a particular instance of an experiment.

\begin{figure*}
    \centering
    \includegraphics[width=0.8\textwidth]{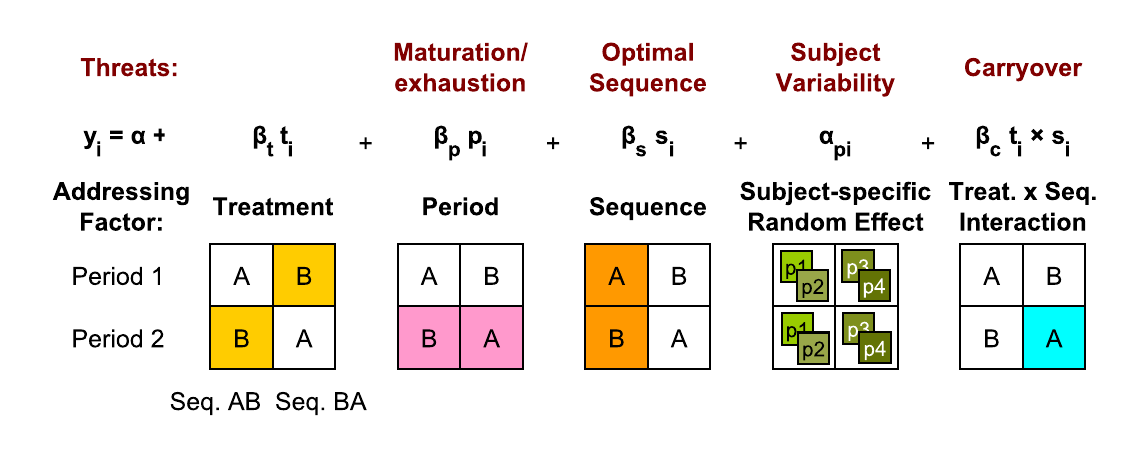}
    \caption{Relevant Factors Influencing the Response Variable in a Crossover-Design Experiment}
    \label{fig:threats}
\end{figure*}

\Cref{fig:threats} visualizes an LMM (second row) that determines the effect of treatment $t_i$ on a response variable $y_i$ while also addressing the above-mentioned threats (first row) by modeling factors that address those threats (third row).
The bottom row visualizes which parts of the AB/BA crossover-design experiment these terms affect.
For example, the treatment $t_i$ affects period 2 in sequence 1 and period 1 in sequence 2 (marked yellow in \Cref{fig:threats}) with strength $\beta_t$.
For simplicity, we constrain the visualization to a crossover-design experiment with one main factor containing two levels (A and B), which requires two periods and two sequences (AB and BA) and is commonly referred to as AB/BA crossover design.

\subsection{State of Practice}
\label{sec:related:previous}

Since the introduction of evidence-based SE by Kitchenham et al.~\cite{kitchenham2004evidence}, the field has been subject to reviews about the state of practice of various aspects of empirical research.
For example, Kampenes et al. reviewed the state of practice of designing, conducting, and evaluating quasi-experiments~\cite{kampenes2009systematic}.
They conclude that terminological ambiguity and a lack of awareness of specific biases affect the validity of drawn conclusions.
Menzies et al. reviewed data analysis practices in empirical SE research~\cite{menzies2019bad} and consolidated 12 ``bad smells'' commonly committed in publications.
Hannay et al. surveyed SE experiments regarding the degree of realism of the employed experimental material and task~\cite{hannay2008role} and detected a lack of awareness of the implications of realism in SE publications.

The aforementioned work by Vegas et al. also contained a secondary study of crossover-design experiments~\cite{vegas2015crossover}, which motivated their guidelines described in \Cref{sec:related:design}.
The review identified similar problems in the state of practice.
Several publications misuse terminology, remain unclear in their design decision, and apply an incorrect analysis to the data from crossover-design experiments.
These results are supported by a similar study by Kitchenham et al. which focused on families of experiments~\cite{kitchenham2019problems}.

The guidelines by Vegas et al.~\cite{vegas2015crossover} have been extended in several regards.
For example, Madeyski et al. investigated and demonstrated the calculation of effect sizes for crossover-design experiments~\cite{madeyski2018effect}.
Kitchenham et al. add the importance of determining and interpreting the correlation between participants' response variable measures~\cite{kitchenham2021importance}.
Cruz et al. propose the use of generalized estimating equations (GEEs) over LMMs to analyze crossover-design data~\cite{cruz2023semiparametric,cruz2024correlation}.
Still, none of these studies has reflected on the impact that the original guidelines~\cite{vegas2015crossover} had on the SE literature, which we aim to contribute in this study.

\section{Method}
\label{sec:method}

To answer our RQ, we conducted a literature survey in the following three steps.
First, we selected an appropriate sample of primary studies (\Cref{sec:method:inclusion}) to be considered in further steps.
Then, we extracted relevant data from these primary studies (\Cref{sec:method:extraction}).
Finally, we analyzed the extracted data (\Cref{sec:method:analysis}).

\subsection{Study Inclusion}
\label{sec:method:inclusion}

An answer to our RQ requires considering empirical SE publications analyzing crossover-design experiments.
We limit this population of primary studies to publications that explicitly cite the guidelines by Vegas et al.~\cite{vegas2015crossover} for two reasons:

\begin{enumerate}
    \item By explicitly evaluating papers citing the guidelines we provide a reflection on the guidelines' impact.
    \item We assume that the proportion of papers correctly analyzing crossover-design experiments is larger in the subset that cites the guidelines than in the subset that does not. We assume that the former provides an upper bound for the proportion of correctly analyzed crossover-design experiments.
\end{enumerate}

We gathered studies by selecting all entries from Google Scholar citing the guidelines~\cite{vegas2015crossover}.
The obtained sample consisted of 136 entries on the 1\textsuperscript{st} of March 2024.
We filtered this sample using the inclusion and exclusion criteria in \Cref{tab:inclusion:criteria}.
We considered a study eligible if it meets all inclusion and none of the exclusion criteria.

\begin{table}
    \centering
    \caption{Inclusion (In) and exclusion (Ex) criteria}
    \label{tab:inclusion:criteria}
    \begin{tabularx}{\columnwidth}{l|X}
        \toprule
        \textbf{ID} & \textbf{Criterion} \\ \midrule
        In1 & The article is related to software engineering. \\
        In2 & The article contains an empirical study as a main contribution. \\
        In3 & The empirical study is an experiment comparing at least two levels (e.g., baseline and treatment) of a main factor. \\
        In4 & The experiment utilizes a crossover design in which all subjects are administered all levels of the main factor. \\
        In5 & The subjects of the experiment are humans. \\ 
        \hline
        Ex1 & The article is not available through the university's access program. \\
        Ex2 & The article is not written in English. \\
        Ex3 & The article is extended by or a duplicate of an already included article. \\
        Ex4 & The article is not peer-reviewed (e.g., a thesis or blog post). \\
        \bottomrule
    \end{tabularx}
\end{table}

Inclusion criteria In1-In4 ensure that the study fits our research goal.
In5 further limits eligible studies to those where the specific benefit of crossover-design experiments (controlling subject variability) is relevant.
Ex1 and Ex2 exclude inaccessible studies, Ex3 removes duplicates, and Ex4 serves as a quality assurance measure.

The first and second authors of this study conducted the inclusion phase.
The 136 articles were distributed among the two authors based on their availability (90 for the first author, 46 for the second).
For each paper, the assigned author read the abstract, introduction, and method section to decide the inclusion and exclusion criteria. 
Only criterion Ex3 (i.e., filtering out duplicates or extensions) was performed centrally by the first author by clustering the sample by author names and investigating similar candidates.
During the inclusion phase, unclear decisions were flagged and later reviewed by the third author of this study.
The third author acted as an arbiter and decided on the three unclear cases.
The inclusion phase identified 48 primary studies ($48/136=35.3\%$) as eligible for the subsequent data extraction phase.

Before conducting the inclusion phase, we assessed the reliability of the criteria by randomly selecting a subset of 14 studies ($14/136=10.3\%$ of the sample) to be rated by both the first and second authors.
The two authors reached a perfect agreement on all ratings, supporting the mutual understanding and reliability of the criteria before proceeding with the main inclusion phase.

\subsection{Data Extraction}
\label{sec:method:extraction}

From each of the 48 eligible primary studies, we extract the attributes in \Cref{tab:extraction} for every individual experiment reported in the study (as one study may conduct and report multiple experiments).

\begin{table}
    \centering
    \caption{Data Extraction Attributes}
    \label{tab:extraction}
    \begin{tabularx}{\columnwidth}{p{1.6cm}|X|l}
        \toprule
        \textbf{Attribute} & \textbf{Description} & \textbf{Type} \\ 
        \midrule
        \multicolumn{3}{l}{\textbf{Subjects}} \\ 
        \hline
        Subject number & Number of human participants & Count \\
        Subject type & Type of participants & Enum \\
        \hline
        \multicolumn{3}{l}{\textbf{Analysis}} \\
        \hline
        Analysis Method & Statistical method applied for inferential analysis of the treatment's effect & Enum \\
        Test Type & (only if the analysis method is NHST) Type of statistical significance test & Enum \\
        Threat Addressal & Way of dealing with the specific threats of validity of crossover-design experiments at analysis time & Enum \\
        Washout & Whether a washout period was scheduled between experimental periods & Bool \\
        \hline
        \multicolumn{3}{l}{\textbf{Material}} \\
        \hline
        Availability & Degree to which material (data set and analysis scripts) are available & Enum \\
        Location & URL of the material if available & URL \\
        \bottomrule
    \end{tabularx}
\end{table}

The attribute group \textbf{subjects} characterizes the number and types of participants involved in each experiment.
The attribute group \textbf{analysis} contains the main variables of interest.
The \textit{analysis method} represents the type of statistical method applied for inference, for example, NHSTs, (generalized) linear models (GLMs), or (generalized) linear mixed models (GLMMs).\footnote{We do not distinguish between GLMs and LMs, nor between GLMMs and LMMs, as only their shared property of containing a random effect or not is relevant to our study.}
If a primary study reported analyzing the data from the experiment using an NHST, we additionally recorded the \textit{test type} (e.g., paired or unpaired T-test, Mann-Whitney U test, etc.).
The attribute \textit{threat addressal} represents the main attribute of interest in the scope of this study.
For each of the four types of threats to validity as mentioned in \Cref{sec:related} (i.e., maturation/exhaustion, optimal sequence, subject variability, and carryover), we determined how the primary study addresses it on the categorical scale shown in \Cref{tab:addressal}.

\begin{table}
    \centering
    \caption{Types of Threat Addressal}
    \label{tab:addressal}
    \begin{tabularx}{\columnwidth}{l|X}
        \toprule
        \textbf{Type} & \textbf{Description} \\
        \midrule
        Modeled & The authors address the threat to validity by modeling the factor in the analysis (e.g., as a parameter in a GLM or GLMM). \\
        Isolated & The authors analyze the threat to validity in isolation, i.e., conduct a statistical test with the threat variable as the only independent variable. \\
        Acknowledged & The authors do not address the threat in the analysis but acknowledge its (unaddressed) influence in the threats to validity section. \\
        Neglected & The authors do not address the threat to validity in the analysis, but claim it is negligible due to the employed design. \\
        Ignored & The authors neither address nor acknowledge the threat to validity. \\
        \bottomrule
    \end{tabularx}
\end{table}

Finally, the attributes from the \textbf{material} group record to what degree both the data obtained by the experiment and the script(s) used to perform the analysis are available. 
We recorded the \textit{availability} attribute based on a previously established, categorical scale of research artifact availability~\cite{frattini2023let} which includes levels like \textit{archived}, \textit{reachable}, and \textit{unavailable}.
If the material was available, we also recorded how to access it in the \textit{location} attribute.

We summarized the extraction guidelines containing a definition, concise extraction rules, as well as examples for each attribute in a shared document. 
To assess the reliability of these guidelines, the first and second authors performed an overlap of the extraction task prior to the main extraction phase.
During this overlap, the two authors applied the extraction guidelines to the seven primary studies from the 14 that were already involved in the inclusion overlap and which were included according to our criteria.
We assessed the agreement of all attributes from the subjects, analysis, and material group except the \textit{location} attribute as it bears no empirical value.
We calculated Pearson's correlation coefficient (PCC) for the numerical attribute \textit{subject number} and Bennett's S-score~\cite{bennett1954communications} between the two ratings of categorical attributes.
None of the seven primary studies in our subset applied a washout period, which is why we did not calculate the inter-rater agreement for this variable.
We selected Bennett's S-score over the more common Cohen's Kappa as the latter is known to be unreliable for uneven marginal distributions~\cite{feng2015mistakes}.
The two raters were in perfect agreement about the \textit{subject number} attribute ($PCC=1.0$) and achieved an average S-score of 85.8\% over all categorical columns.
After confirming the reliability of the extraction guidelines, the two first authors proceeded with applying them to the 48 primary studies originally assigned to them during the inclusion phase.
The third author acted as an arbiter and clarified seven unclear instances in the data extraction.

\subsection{Data Analysis}
\label{sec:method:analysis}

Upon completion of the data extraction phase, we generated descriptive statistics from the distribution of attribute values.
We visualized categorical data using bar charts and numerical data using box plots.
We represented the main attribute of interest, the \textit{threat addressal}, as a heatmap where one axis listed the four threats to validity and the other axis the types of threat addressal.

\section{Results}
\label{sec:results}

The 48 primary studies describe 67 experiments.
Main factors of interest include, among others, testing paradigms like test-driven development~\cite{baldassarre2022affective,esposito2023test,fucci2016external}, the effect of API design rules~\cite{bogner2023restful}, and noise during software development~\cite{romano2020effect}.
The following subsections report the obtained results grouped by the attribute groups in \Cref{tab:extraction}.

\subsection{Subjects}
\label{sec:results:subjects}

\begin{figure}
    \centering
    \includegraphics[width=0.8\columnwidth]{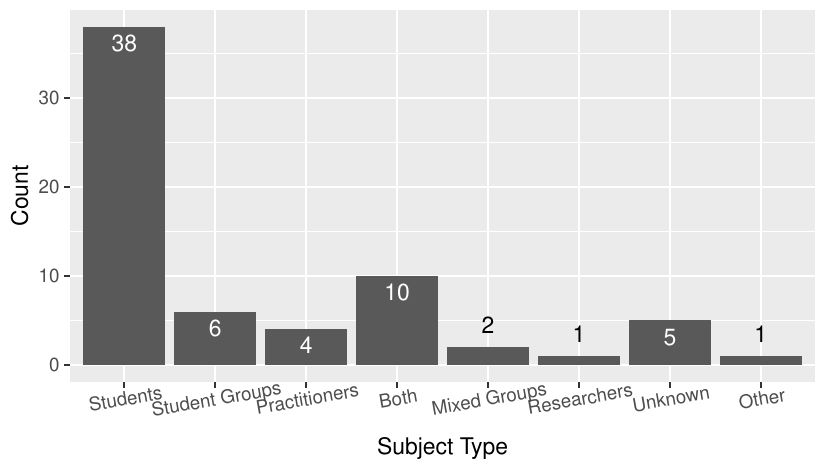}
    \caption{Types of subjects in the experiments}
    \label{fig:subject:type}
\end{figure}

\Cref{fig:subject:type} visualizes the distribution of subject types among the experiments.
The predominant type of participants are students, while practitioners are underrepresented.
In five cases, the authors do not state the participant type at all.
The case labeled as ``other'' sampled from app users, which were only partly students~\cite{francese2020minijava}.

\begin{figure}
    \centering
    \includegraphics[width=0.8\columnwidth]{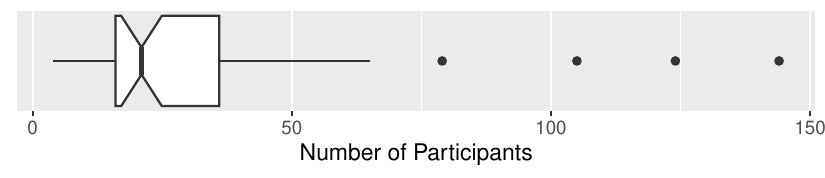}
    \caption{Number of subjects in the experiments}
    \label{fig:subject:count}
\end{figure}

\Cref{fig:subject:count} visualizes the distribution of subject count among the experiments (median of 21, mean of 31.2).
Notable outliers are experiments with 144~\cite{coppola2023effectiveness}, 124~\cite{khan2019aspectocl}, and 105~\cite{bogner2023restful} participants.
The former two involved students, the latter both students and practitioners.
One study did not mention the number of participants involved in the experiment~\cite{trkman2019impact}.

\subsection{Analysis}
\label{sec:results:analysis}

\begin{figure}
    \centering
    \includegraphics[width=0.8\columnwidth]{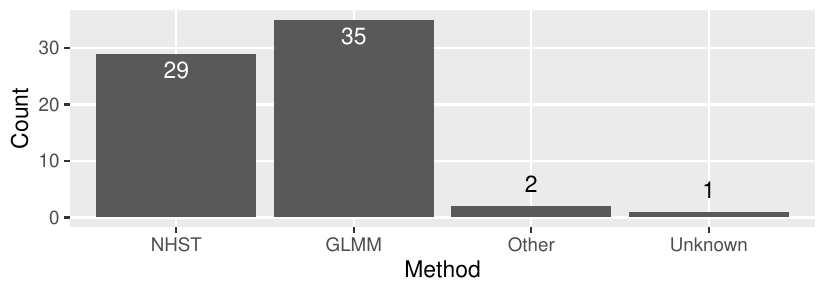}
    \caption{Applied statistical methods}
    \label{fig:analysis:method}
\end{figure}

\begin{figure}
    \centering
    \includegraphics[width=0.8\columnwidth]{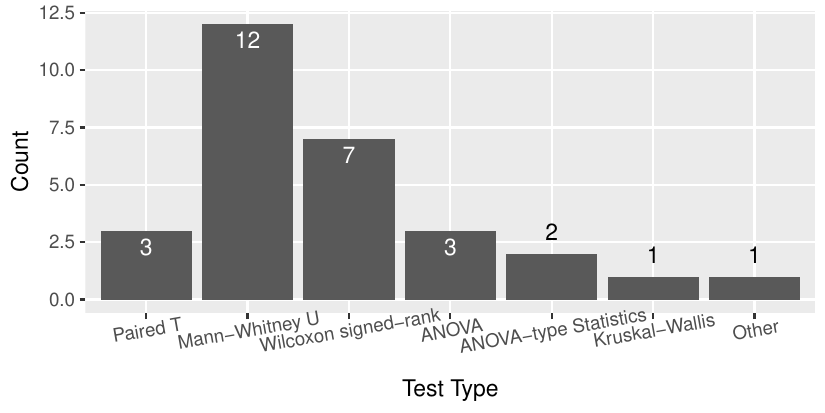}
    \caption{Applied NHSTs}
    \label{fig:analysis:nhst}
\end{figure}

\Cref{fig:analysis:method} visualizes the distribution of applied statistical methods.
The applied methods are largely limited to NHSTs and LMMs.
Exceptions (coded as ``other'') are papers that, for example, only compare the mean values of the response variable stratified by the treatments~\cite{schneid2022semi}.
\Cref{fig:analysis:nhst} shows the distribution of test types applied to the experiments that analyzed their data using an NHST (n=29).

\begin{figure}
    \centering
    \includegraphics[width=\columnwidth]{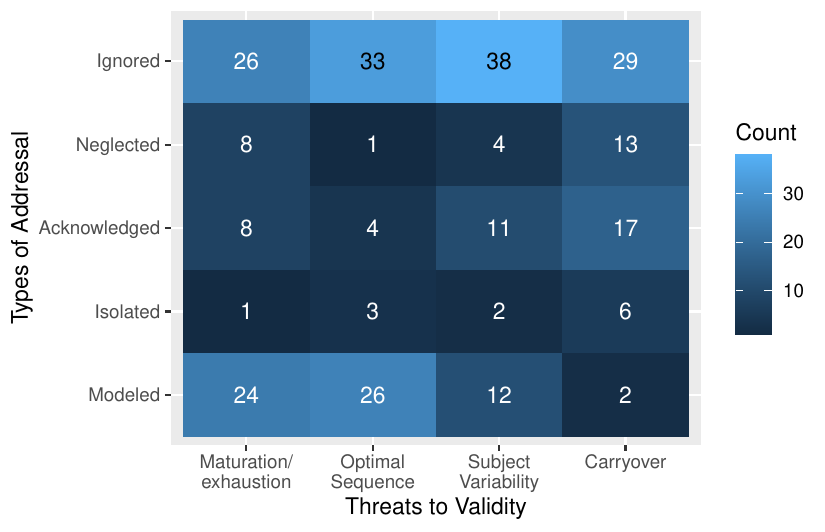}
    \caption{Addressal of the Threats to Validity}
    \label{fig:analysis:addressal}
\end{figure}

\Cref{fig:analysis:addressal} shows how authors address the threats to validity detailed by Vegas et al.~\cite{vegas2015crossover} in their analysis, i.e., the figure visualizes the distribution of types of addressal per threat to validity.
For example, the first, top-left cell indicates that for 26 experiments, the maturation/exhaustion threat was ignored.
The visualization shows that the maturation/exhaustion and the optimal sequence threat are mostly either ignored completely or modeled via the period and sequence variable respectively.
The subject variability and carryover threat are mostly either ignored or acknowledged.
Rarely do authors analyze the threat type in isolation (4.5\% of all cases).
In total, 47.0\% of all threats to validity ($(26+33+38+29)/(4\times67)$) were simply ignored by the primary studies in our sample.
A subset of papers at least acknowledges these threats, but they either leave it at this acknowledgment (14.9\% of threats are \textit{acknowledged}) or claim that the threat is mitigated by design (9.7\% of threats are \textit{neglected}).
The discouraged addressal in isolation only occurs rarely (4.5\% of threats are \textit{isolated}) while only 23.8\% of the threats to validity are explicitly modeled.
None of the papers in our sample contained an analysis that follows the guidelines completely, i.e., addressed all four threats to validity by modeling the respective factors.
The experiment where the analysis comes closest to the recommended guidelines was performed and reported by B\"under et al.~\cite{bunder2019towards}, who \textit{modeled} the period, sequence, and subject variability in their GLMM and \textit{acknowledged} the carryover threat.
However, this was still considered valid by the original guidelines, especially when the carryover threat is confounded with other treats~\cite{vegas2015crossover}.

Only three experiments include a washout period~\cite{moreno2024software,baldassarre2022affective,romano2020effect}.
The duration of the washout periods varies between 30 minutes~\cite{romano2020effect} and a day~\cite{baldassarre2022affective} or was not specified~\cite{moreno2024software}.

\subsection{Material}
\label{sec:results:material}

\begin{figure}
    \centering
    \includegraphics[width=0.8\columnwidth]{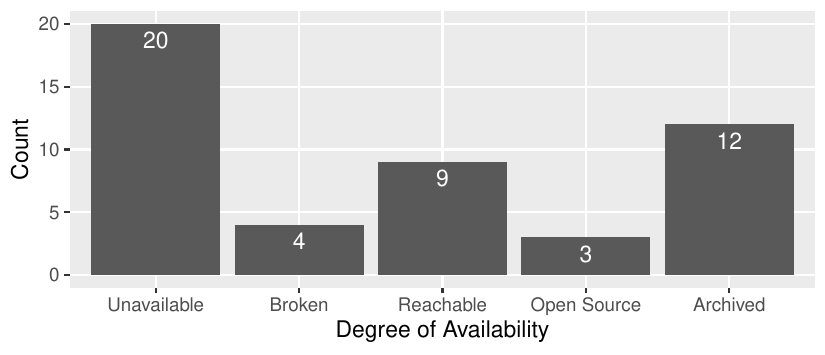}
    \caption{Availability of Data Sets}
    \label{fig:material:data}
\end{figure}

\begin{figure}
    \centering
    \includegraphics[width=0.8\columnwidth]{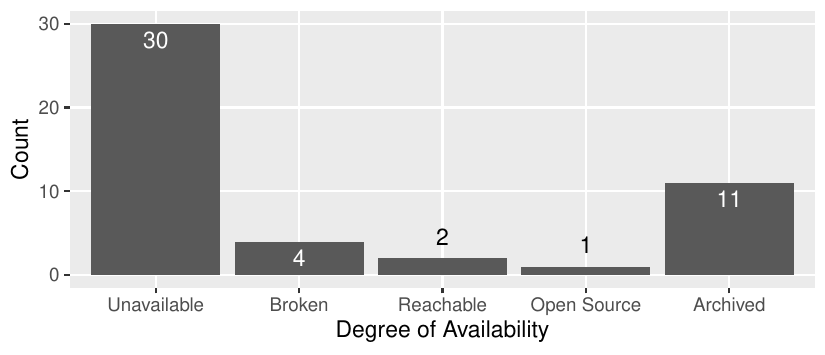}
    \caption{Availability of Analysis Scripts}
    \label{fig:material:analysis}
    \vspace{-0.3cm}
\end{figure}

\Cref{fig:material:data,fig:material:analysis} visualize the availability of data sets and analysis script.\footnote{Note that these are distributions among the 48 publications, not among the 67 experiments, because, in our sample, we observed that any additional material was always associated with a publication, not an individual experiment.}
Most of the material was never available (20 data sets and 30 scripts are \textit{unavailable}) and a portion has become unavailable since their original publication (4 data sets and 4 scripts are \textit{broken}).
Among the remaining material, several data sets ($12/48=25\%$) and scripts ($12/48=22.9\%$) have been properly \textit{archived} and, therefore, preserved for future use like reproduction~\cite{frattini2024requirements}.

\section{Discussion}
\label{sec:discussion}

Despite clear guidelines~\cite{vegas2015crossover}, the majority of the subset of SE papers that indicate that they are aware of them through citation do not comply with them.
This indicates that SE papers reporting the analysis of crossover-design experiments run the risk of drawing incorrect conclusions due to their incomplete analysis~\cite{kitchenham2003case,vegas2015crossover}.
The portion of experiments where a threat to validity was neglected shows that some authors assume that the crossover design mitigates this threat by default.
However, the crossover design merely counterbalances, i.e., de-confounds the threats from the effect of the treatment, but the effect still applies to the response variable and needs to be modeled when analyzing the data.

The results imply that the guidelines~\cite{vegas2015crossover} did not unfold the full effect that the authors hoped to achieve with their contribution.
Many experiments published in SE literature claim to follow established guidelines but fail to do so.
A possible reason for this is that the guidelines were not supplemented with the analysis scripts that could have provided practical guidance on how to implement them.

However, despite the remaining room for improvement, we observed positive effects of the guidelines.
While the original literature survey reported that none of the 38 primary studies in its sample dealt with carryover at analysis time~\cite{vegas2015crossover}, our sample showed at least two studies (3\%) that modeled the carryover effect and six that analyzed it in isolation (9\%).
17 (25.3\%) at least acknowledged the carryover threat.
Additionally, while the original literature survey observed only one primary study that explicitly defined its experimental periods, exactly half of our sample (24 papers) contained either a table~\cite{bunder2019towards,francese2020minijava,iniguez2020improvement} or figure~\cite{fucci2018longitudinal,romano2020effect} visualizing the periods and sequences of their experimental design.

To improve future analyses of crossover-design experiments in SE, we recommend increasing the awareness of established guidelines~\cite{vegas2015crossover}.
Including these guidelines in textbooks~\cite{wohlin2012experimentation} and standards~\cite{ralph2020empirical} will aid authors in more rigorous analyses.
Furthermore, we urge reviewers to put more emphasis on guideline adherence.
This not only requires awareness of the existence of guidelines (i.e., checking that appropriate guidelines were cited) but also of their content (i.e., checking that the guidelines were properly followed).

\section{Limitations and Future Work}
\label{sec:limit}

\subsection{Threats to Validity}
\label{sec:limit:threats}

Our study suffers from the following threats to validity~\cite{wohlin2012experimentation}.
Most prominently, our study is subject to a threat to external validity.
While we aim to draw general conclusions about experimentation in SE literature, our sample is limited to SE literature that cites the investigated guidelines~\cite{vegas2015crossover}.
However, we argue that our sample is adequate for this paper for the following reasons.
The guidelines~\cite{vegas2015crossover} are---to the best of our knowledge---the only SE-specific guidelines for analyzing crossover-design experiments in SE.
Hence, authors of our selected subset had access to the only guidelines explicitly advising them on how to properly analyze their experimental data.
Therefore, we are confident that our sample represents an upper bound to extrapolate to the SE experimentation literature.

Additionally, our study inclusion and data extraction phases were subject to two threats to construct validity.
Firstly, both phases involved subjective judgment of criteria and extraction guidelines.
We mitigated this threat by quantifying the inter-rater agreement of both phases.
Given the satisfactory inter-rater agreement, we are confident in the reliability of our results.
Secondly, the categories of the \textit{threat addressal} attribute were devised ad-hoc and not based on an existing taxonomy. 
We addressed this threat through thorough discussions among all three authors.

\vspace{-0.2cm}

\subsection{Future Work}
\label{sec:limit:limitations}

In the scope of this study, we only surveyed how authors addressed the four threats to validity presented in \Cref{sec:related:design} as per the original guidelines~\cite{vegas2015crossover}.
We did not consider additional threats to validity which crossover-design experiments are subject to.
Additional threats we plan to investigate is the \textit{material} threat, i.e., the influence of the used experimental material (e.g., code snippet), and the interaction effects between subjects and treatments, e.g.,, personal preference for specific treatments.

Furthermore, we plan to extend our sample by including SE studies presenting experiments with a crossover design that do not cite the guidelines by Vegas et al.~\cite{vegas2015crossover}.
By comparing this sample with our current one we aim to further characterize the impact of these guidelines.

Finally, we plan to reproduce the analysis of the surveyed experiments as far as possible.
While we cannot reproduce the analysis of experiments where the raw data is not available, we plan to (1) reproduce the original analyses as described in the publication, and (2) reanalyze the data according to the data analysis guidelines~\cite{vegas2015crossover}.
Contrasting these analyses will produce more examples for an application of the guidelines, but also reveal studies that drew incorrect conclusions due to incorrect data analyses.
This quantifies the effect of the guidelines in terms of the risk of drawing inappropriate conclusions and will serve as further motivation for adherence.
Additionally, we plan to analyze and reproduce crossover-design experiments that do not cite the data analysis guidelines to compare them with our current sample and, therefore, further study the guidelines' impact.

\section{Conclusion}
\label{sec:conclusion}

In this reflection, we investigate the state of practice of analyzing crossover-design experiments in SE.
A sample of publications citing explicit guidelines~\cite{vegas2015crossover} shows that the state of practice still contains several, significant gaps threatening the validity of the drawn conclusions.
While the guidelines by Vegas et al.~\cite{vegas2015crossover} supported a significant portion of authors to analyze their obtained data at least in parts, the general sample of papers shows room for improvement.

We hope that the overview of the state of practice of analyzing crossover-design experiments in SE encourages authors to investigate this topic more thoroughly and that both our visualizations as well as the recovery of the data analysis script from the original guidelines~\cite{vegas2015crossover} will support the latter in guiding authors towards a correct analysis.
We encourage authors to abandon the overly simple NHST analysis~\cite{frattini2024second} for more complex (G)LMMs, which enable them to adequately address threats to the validity of crossover-design experiments during analysis.

\begin{acks}
    This work was supported by the KKS foundation through the S.E.R.T. Research Profile project at Blekinge Institute of Technology and partially supported by grant PID2022-137846NB-I00 funded by MCIN/AEI/10.13039/501100011033 and by \say{ERDF A way of making Europe}.
\end{acks}

\bibliographystyle{ACM-Reference-Format}
\bibliography{material/references}

\end{document}